# Low Power, Scalable Nanofabrication *via* Photon Upconversion

Qi Zhou[1], Hao-Chi Yen[1,2], Qizhen Lan[1,2], Arynn O. Gallegos[1], Manchen Hu[1], Kyle Frohna[1,2], Hannah Niese[1,3], Da Lin[1,2], Natalia Murrietta[1], Pournima Narayanan[1,4], Tracy H. Schloemer[1], Linda Pucurimay[1,2], Sebastian Fernández[1], Michael Seitz[1,5], Daniel N. Congreve[1,*]

[1]Department of Electrical Engineering, Stanford University, Stanford, CA, United States

[2]Department of Materials Science and Engineering, Stanford University, Stanford, CA, United States

[3] Current Address: Department of Mechanical and Process Engineering, ETH Zurich, 8092 Zurich, Switzerland

[4]Department of Chemistry, Stanford University, Stanford, CA, United States

[5]Condensed Matter Physics Center (IFIMAC) and Department of Condensed Matter Physics, Autonomous University of Madrid, Madrid, Spain

[*]e-mail: congreve@stanford.edu

# Abstract

Micro- and nanoscale fabrication, which enables precise construction of intricate three-dimensional structures, is of foundational importance for advancing innovation in plasmonics, nanophotonics, and biomedical applications. However, scaling fabrication to industrially relevant levels remains a significant challenge. We demonstrate that triplet-triplet annihilation upconversion (TTA-UC) offers a unique opportunity to increase fabrication speeds and scalability of micro- and nanoscale 3D structures. Due to its nonlinearity and low power requirements, TTA-UC enables localized polymerization with nanoscale resolutions while simultaneously printing millions of voxels per second through optical parallelization using off-the-shelf light-emitting diodes and digital micromirror devices. Our system design and component integration empower fabrication with a minimum lateral feature size down to 230 nm and speeds up to 112 million voxels per second at a power of 7.0 nW per voxel. This combination of high resolution and fast print speed demonstrates that TTA-UC is a significant advancement in nanofabrication technique, evidenced by the fabrication of hydrophobic nanostructures on a square-centimeter scale, paving the way for industrial nanomanufacturing.

## Main text

## Introduction and Motivation

The demand for scalable fabrication of complex three-dimensional (3D) structures featuring micro- and nanoscale details is surging across diverse fields, including nano-optics and photonics,[1,2] microfluidics,[3,4] microactuators,[5] and biomedicine.[6,7] Light-based 3D printing stands out as an attractive solution for crafting these intricate structures with high-resolution features. This process typically begins when a photoinitiator molecule absorbs an ultraviolet (UV) or blue photon, generating radical species that trigger a polymerization reaction. However, excitation by a one-photon absorption (1PA) process is linearly proportional to light intensity, often resulting in polymerization along the entire light path and producing features larger than the diffraction limit (**Fig. 1A**). In contrast, two-photon polymerization (2PP),[8–11] an established technique for creating nanoscale objects with arbitrary architectures, leverages a nonlinear optical process known as two-photon absorption (2PA). Unlike 1PA, 2PA process exhibits a quadratic dependence on light intensity. This leads to a threshold behavior, enabling precise spatial control of the polymerization at the focal point of the incident light down to nanoscale resolution in both lateral and vertical dimensions (**Fig. 1B**). Nevertheless, 2PA is an inherently inefficient process as it involves a virtual molecular state with extremely short lifetimes,[11] necessitating high-powered femtosecond pulsed lasers, increasing both cost and complexity while limiting printing speeds.

2PP is considered the gold standard with regard to resolution and accuracy. Commercially available products are widely used on the research scale.[1,3,5,7] Yet moving towards industrial-scale manufacturing remains a significant hurdle as a single volumetric element (voxel) traces out the 3D structure. Parallelizing the 2PP process to simultaneously print many voxels and increase printing speeds remains a major challenge. While 2PP parallelization has been demonstrated,[12–14] the input power demand increases with the number of foci, making industrial-scale production of 3D nanostructures difficult. Additionally, excessive light dosage can cause resin boiling or damage to the photopolymer, significantly hindering printing speed and reliability.[15,16]

To circumvent these challenges, researchers have demonstrated promising micro- and nanoprinting technologies aimed at enabling more scalable fabrication utilizing relatively low-powered continuous-wave (CW) lasers, including dual-color polymerization[17] and light-sheet 3D laser microprinting.[18–20] While these exciting techniques have displayed excellent print fidelity and speed, the achievable vertical resolution is directly dependent on the width of the light sheet,[17] which can limit the ability to achieve nanoscale resolution in the vertical direction, and require orthogonal excitation for the light sheet.

Triplet–triplet annihilation upconversion (TTA-UC) is another intriguing nonlinear process that has drawn researchers' attention for 3D printing due to its low required input power and quadratic nature.[21–24] TTA-UC converts two low-energy photons into one higher-energy photon by manipulating molecular excitonic states. The detailed mechanism is illustrated in **Fig. 1C**. Importantly, the initial photon absorption is linear (i.e., only one photon is absorbed per sensitizer excitation) but the overall process exhibits quadratic dependence on light intensity, as two triplet species must annihilate. TTA-UC simultaneously provides the quadratic nature of 2PA with the low input power requirements of 1PA, thanks to the long-lived nature of the triplet states.[25] When applied to 3D fabrication, TTA-UC allows for nanoscale resolution while enabling large-scale parallelization. As a proof of concept, we previously developed a strategy for 3D single-focus scanning and 2D parallelized printing using only a few milliwatts of input power.[26] Limberg et al.,[27] Awwad et al.,[28] and Wang et al.[29] further demonstrated the feasibility of TTA for enabling microfabrication with single focus 3D scanning or 2D microfabrication with photomask. While these proof-of-principle reports represent progress, significant work remains to deliver on the promised rapid print speeds of full 3D structures.

In this work, we demonstrate the potential for rapid, scalable 3D nanomanufacturing capabilities. With parallelized TTA-UC printing using light-emitting diodes (LEDs), we produced voxels as small as 0.020 µm$^3$ and printed microstructures over an area of up to 1 cm$^2$ through optimized design of the material, optical, and computational components of the system. In this multi-pronged approach, we address key challenges in all three domains, suppressing undesired parasitic initiation, designing controlled photon delivery, and enabling large-scale parallelization of the optical field. Finally, we demonstrate the potential for scaling manufacturing beyond R&D, as exemplified by diffraction patterns fabricated with a speed of 4.35 mm$^2$/min and hydrophobic structures constructed with a speed of 0.36 mm$^2$/min over a square-centimeter area. Compared to state-of-the-art fabrication technologies, our approach maintains minimum feature sizes similar to 2PP while achieving significantly lower power requirements, higher printing speeds, and enhanced scalability.

# Results

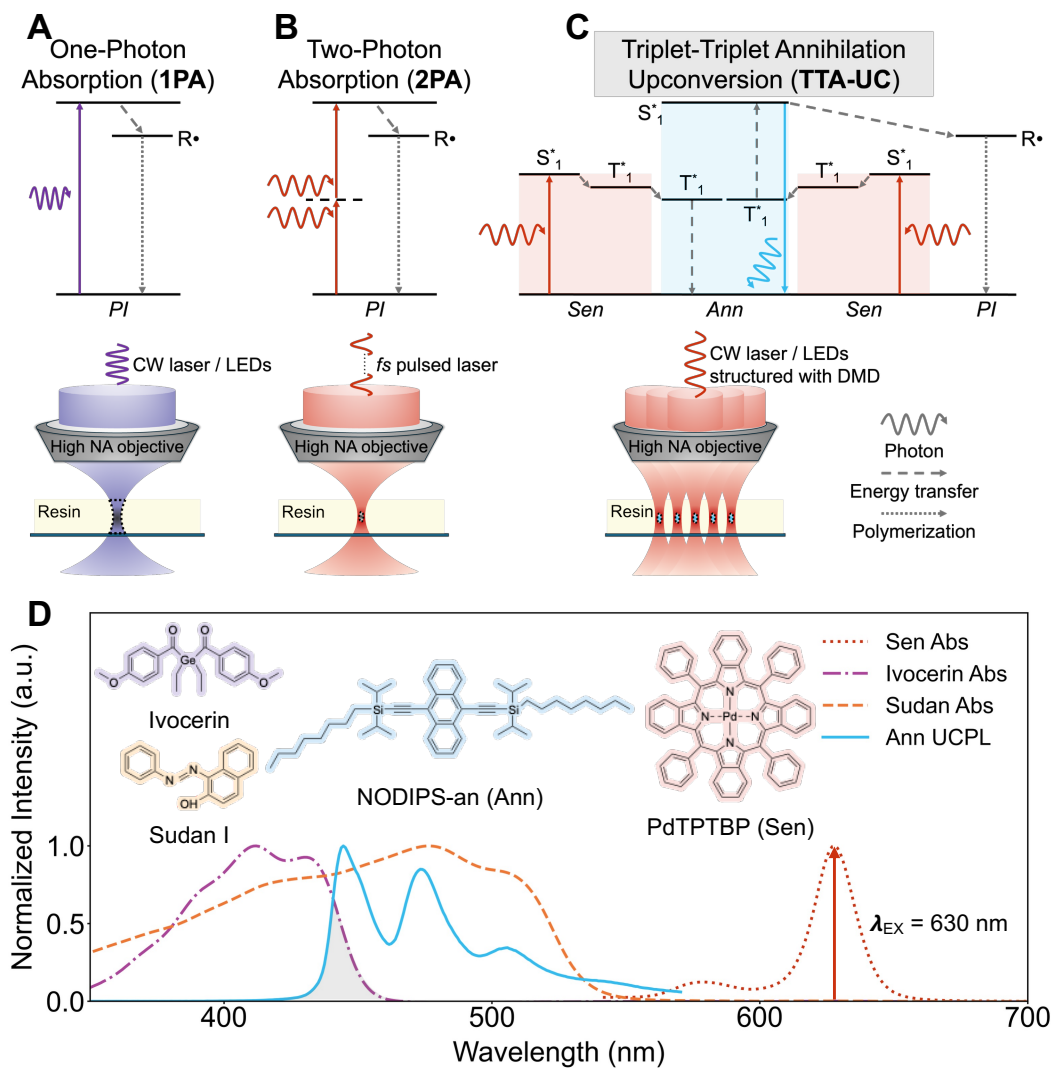

**Figure 1. Resin design incorporating upconversion. (A - C)** Schematic representations illustrate absorption energy level transitions and the polymerization process initiated by photoinitiator (PI) molecules (top), as well as the polymerization volume under various excitation methods (bottom): (A) one-photon absorption (1PA), (B) two-photon absorption (2PA), and (C) triplet–triplet annihilation upconversion (TTA-UC). In 1PA, a single UV/blue photon is absorbed, and excitation is linearly proportional to light intensity, leading to polymerization along the light path (> diffraction limit) (black dashed outline). In 2PA, two near-infrared photons are absorbed, limiting polymerization to the focal point above a threshold (< diffraction limit) (black dashed outline). In TTA-UC, excitation also exhibits a quadratic dependence on light intensity, enabling voxel formation beyond the diffraction limit similar to 2PA (black dashed outline). Notably, TTA-UC requires significantly lower input power, allowing for parallelized polymerization. To achieve this UC process, sensitizer molecules absorb low-energy photons and generate triplet states via spin–orbit coupling, often aided by the presence of heavy metal atoms. These triplet states are then transferred to annihilator molecules through triplet energy transfer (TET). Two annihilator triplet states undergo triplet–triplet annihilation to form one high-energy singlet excited state. This state can either radiatively decay, emitting a photon with higher energy than the incident photons, or transfer energy to a photoinitiator, generating

radicals and initiating polymerization. **(D)** Normalized absorption (Abs) spectra of PdTPTBP (Sen), Ivocerin, and Sudan I, and the UC photoluminescence (PL) spectrum of NODIPS-an (Ann) in toluene, along with their molecular structures. The overlap between Ivocerin absorption and NODIPS-an UCPL is highlighted in gray.

To incorporate TTA-UC materials directly into a 3D printing resin, efficient upconversion in the chemical environment is required to supply energy for polymerization. This necessitates the following: upconverted energy that can be effectively transferred to the photoinitiator (i.e., sufficient spectral overlap between the photoinitiator's absorption and the annihilator's upconversion emission), polymerization initiation that primarily originates from TTA-UC, and polymerization that can be terminated effectively. The latter is especially vital for achieving high vertical resolution.[30]

Prior studies have shown that the sensitizer alone can initiate polymerization even in the absence of a photoinitiator.[27,28] This phenomenon likely originates from the sensitizer's highly efficient triplet generation, which can directly transfer to photoinitiators or monomers and trigger polymerization. Although this unwanted side reaction is reported to occur at a rate several orders of magnitude slower than the upconversion-driven reaction,[27] this 1PA process must still be carefully managed to maintain high resolution and fidelity.

Based on these criteria, the resin components and their respective functions are illustrated in **Fig. S1**. We selected di(trimethylolpropane) tetraacrylate (DTMPTA) as the primary monomer due to its high reactivity and cross-linking ability, and incorporated N,N-dimethylacrylamide (DMAA) to ensure resin phase homogeneity. The absorption spectra of the sensitizer, photoinitiator, and light blocker, as well as the upconversion photoluminescence (PL) of the annihilator, are shown in **Fig. 1D**. As the upconversion pair, we employed palladium(II)-meso-tetraphenyl-tetrabenzoporphyrin (PdTPTBP) and 9,10-bis((diisopropyl(octyl)silyl)ethynyl)anthracene (NODIPS-an) to convert red photons into blue photons, supplying the energy required for polymerization. While 9,10-bis((triisopropylsilyl)ethynyl)anthracene (TIPS-an) is a commonly used annihilator, its poor solubility in DMAA (<1 mg/mL) limited upconversion efficiency in the photoresin. In contrast, by increasing the length of the alkyl chains, NODIPS-an dissolves in DMAA at concentrations exceeding 12.5 mg/mL for improved upconversion performance. With this red-to-blue upconversion system, we selected Ivocerin[26] as the photoinitiator due to its favorable spectral overlap with the UCPL of NODIPS-an, as highlighted in **Fig. 1D**. Additionally, we incorporated 1-phenylazo-2-naphthol (Sudan I) as a light blocker and bis(2,2,6,6-tetramethyl-4-piperidyl-1-oxyl) sebacate (BTPOS) as a carbon radical inhibitor to control both the propagation distance of upconverted light and terminate polymerization reactions, respectively. Through precise control of component concentrations, we significantly reduced the undesired 1PA reaction, making it at least two orders of magnitude slower than the TTA-UC initiation pathway (**Fig. S2**). Interestingly, we also found that the inclusion of Sudan I played a more critical role than initially anticipated. In addition to acting as a light blocker with strong absorption overlap with the UCPL, as shown in **Fig. 1D**, Sudan I may also function as a triplet scavenger, helping to mitigate unwanted side reactions initiated by the sensitizer, thereby containing polymerization more effectively and improving resolution (**Supplementary Note 1, Fig. S3 - 7, Table S1** and **S2**).

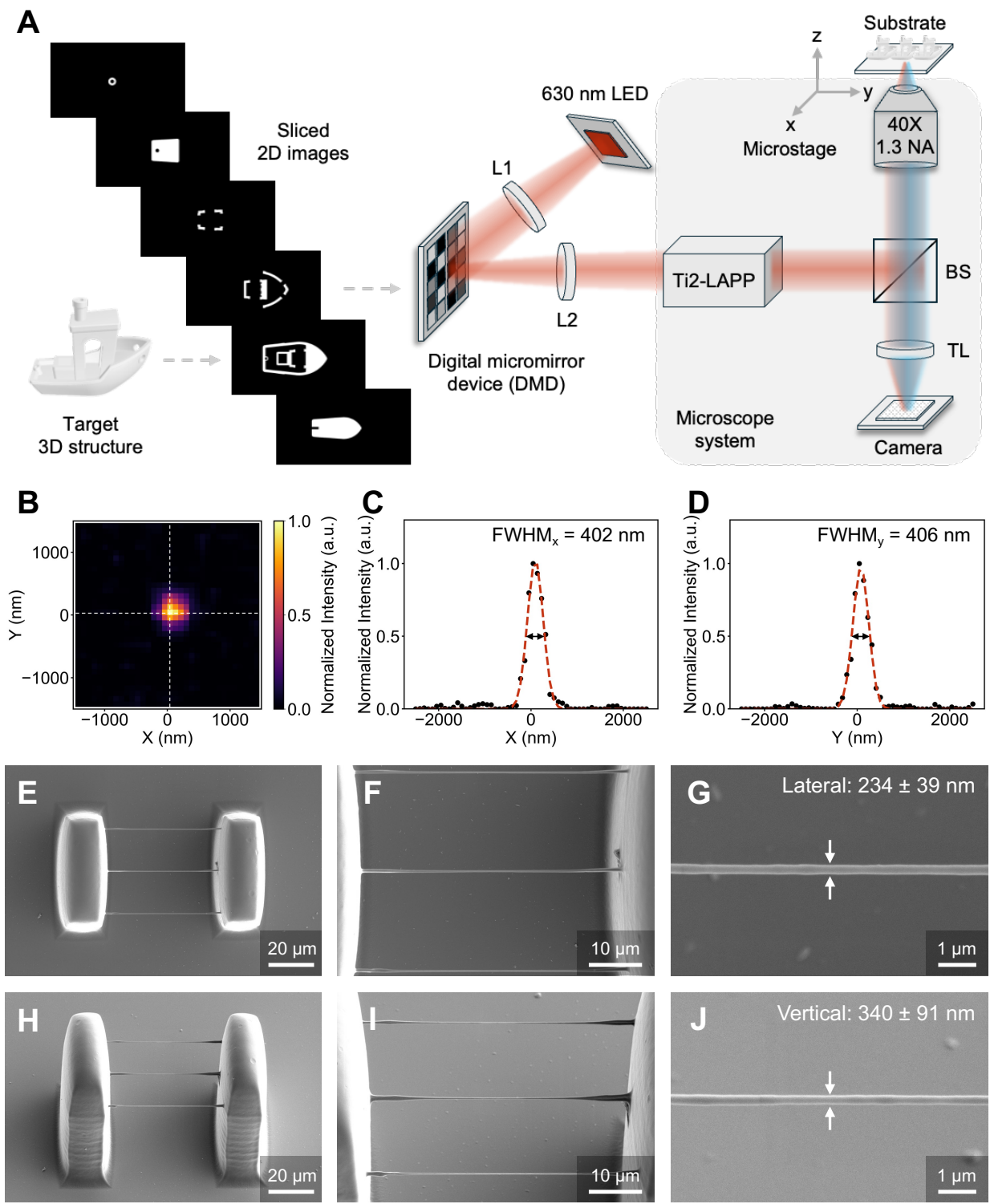

**Figure 2. Precise control of photon delivery. (A)** Schematic diagram of the optical setup and nanofabrication workflow using 3D benchy[31] as an example. L: lens. BS: beam splitter. TL: tube lens. NA: numerical aperture. **(B)** False-colored microscopic image showing scattered light from gold nanoparticles with a diameter of 100 nm. The scattered light represents the point spread function (PSF) of the overall

optical setup. **(C, D)** Gaussian fits of the PSF in the (C) x-direction and (D) y-direction, showing full width at half maximum (FWHM) values of 402 nm and 406 nm, respectively. **(E - G)** SEM images (top-down view) of a three-line bridge structure used to assess the lateral minimum feature size, based on the given resin design and optical setup. **(H - J)** SEM images (tilted view at 40°) of the same three-line bridge shown in (E - G) used to evaluate the vertical minimum feature size Measured values are multiplied by 1.34 to account for the tilt angle. Three different bridges, each containing three distinct lines, with each line measured at three different points, were used to calculate the average values shown in (G) and (J).

We next tested the resin performance using an optics setup integrated with an inverted microscope. As previously mentioned, TTA-UC's low input power requirements allow for straightforward optical parallelization. The optical setup and printing workflow are summarized in **Fig. 2A**. Briefly, we coupled a digital micromirror device (DMD) with an objective lens to project two million pixels simultaneously. First, a collimating lens placed in front of an LED light source ensures uniform light projection onto the DMD (1920 × 1080 pixels, 7.56 µm micromirror pitch). The light reflected from the DMD is then collimated again by a lens and a Nikon Ti2-LAPP system[32] to achieve more even illumination at the back aperture of the objective lens (1.3 numerical aperture (NA), 40×). A microstage, synchronized with the DMD, moves the resin sample holder to print successive 2D slices throughout the 3D volume.

We measured the point spread function (PSF) of the system following the protocol in ref[33] (see Methods section for details) to predict resolution capabilities based on optical hardware alone. The PSF was determined by analyzing the scattering from a 100 nm gold nanoparticle: **Fig. 2B** shows a false-colored image of the scattering, while **Fig. 2C** and **2D** display Gaussian fits for the data highlighted by dashed lines. The full width at half maximum (FWHM) values for the PSF in the x- and y-directions are 402 nm and 406 nm, respectively. Theoretical calculations, based on the objective, the central wavelength of the LED, and the refractive index of the medium, predict a lateral resolution of 296 nm and a vertical resolution of 1133 nm.[34] The measured PSF values are reasonably close to the theoretical lateral resolution. We estimate the vertical resolution to be approximately three to four times the lateral resolution based on Abbe resolution approximations,[34] which implies 1.2 - 1.6 µm for our system. The discrepancy in theoretical and measured resolution may stem from the wavelength selectivity of the DMD and the use of an LED light source with a relatively broad emission range of 630 ± 10 nm.

We then evaluated the minimum feature size achievable by the printing setup. As shown in **Fig. 2E - J**, a three-line bridge structure was designed to test lateral and vertical minimum feature sizes. **Fig. 2E - G** show different magnifications of the bridge from a top-down view captured by scanning electron microscopy (SEM). The line widths indicate a lateral feature size of 234 ± 39 nm (N = 27). **Fig. 2H - J** provide tilted views (40°) of the bridge, revealing a vertical feature size of 340 ± 91 nm (N = 27), after adjustment by a factor of 1.34 to account for the tilting effect. Notably, the vertical feature size is significantly smaller than the optical limit, confirming the importance of TTA-UC's nonlinear properties. We highlight that the three-line projection required edge correction for proximate voxels, as shown in **Fig. S8**. Pre-exposure of the two holders caused radical accumulation, leading to edge expansion of the lines (**Fig. S8**, left). Manual modification of the line image edges significantly reduced this effect, resulting in more uniform lines. This observation suggests that accumulated exposure in 3D fabrication is substantial and requires correction to maintain printing fidelity (*vide infra*).

Based on these results, we calculated the total voxel printing rate following previous procedure,[18] a metric that enables meaningful comparisons across 3D printing approaches with varying voxel volumes. The DMD has a resolution of 1920 × 1080 pixels, equivalent to $2.07 \times 10^6$ pixels or approximately $1.56 \times 10^6$ voxels, assuming a round voxel shape with the measured voxel diameter. With an exposure time of 14 ms, as used for the line printing, we achieve a voxel printing rate of $112 \times 10^6$ voxels/s for a voxel volume of $(0.24 \text{ μm})^2 \times 0.34 \text{ μm} = 0.020 \text{ μm}^3$, highlighting the high printing speed enabled by this parallelization approach.

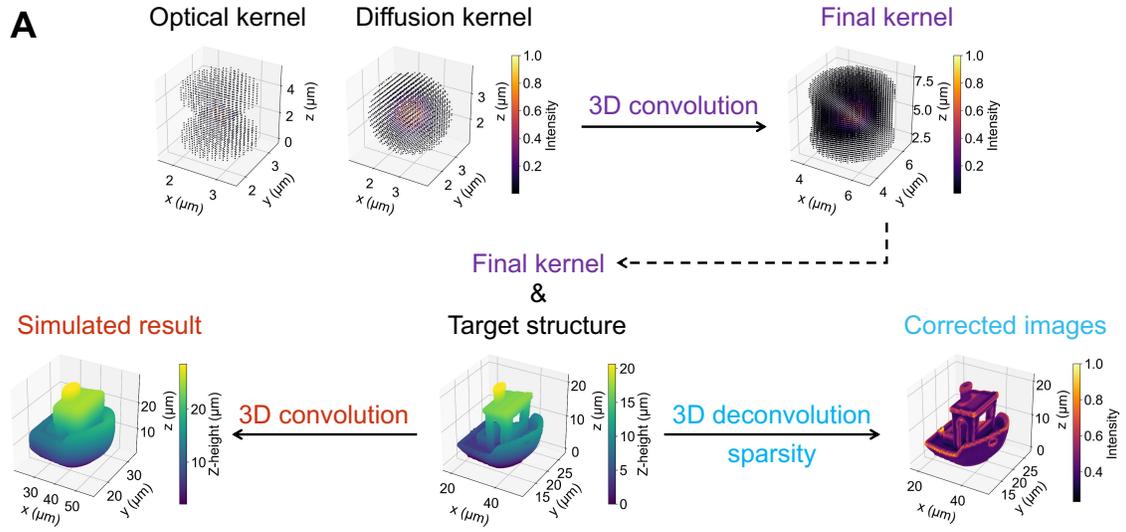
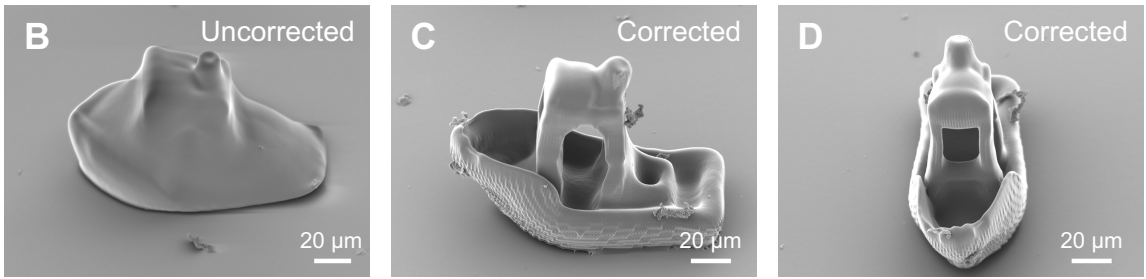
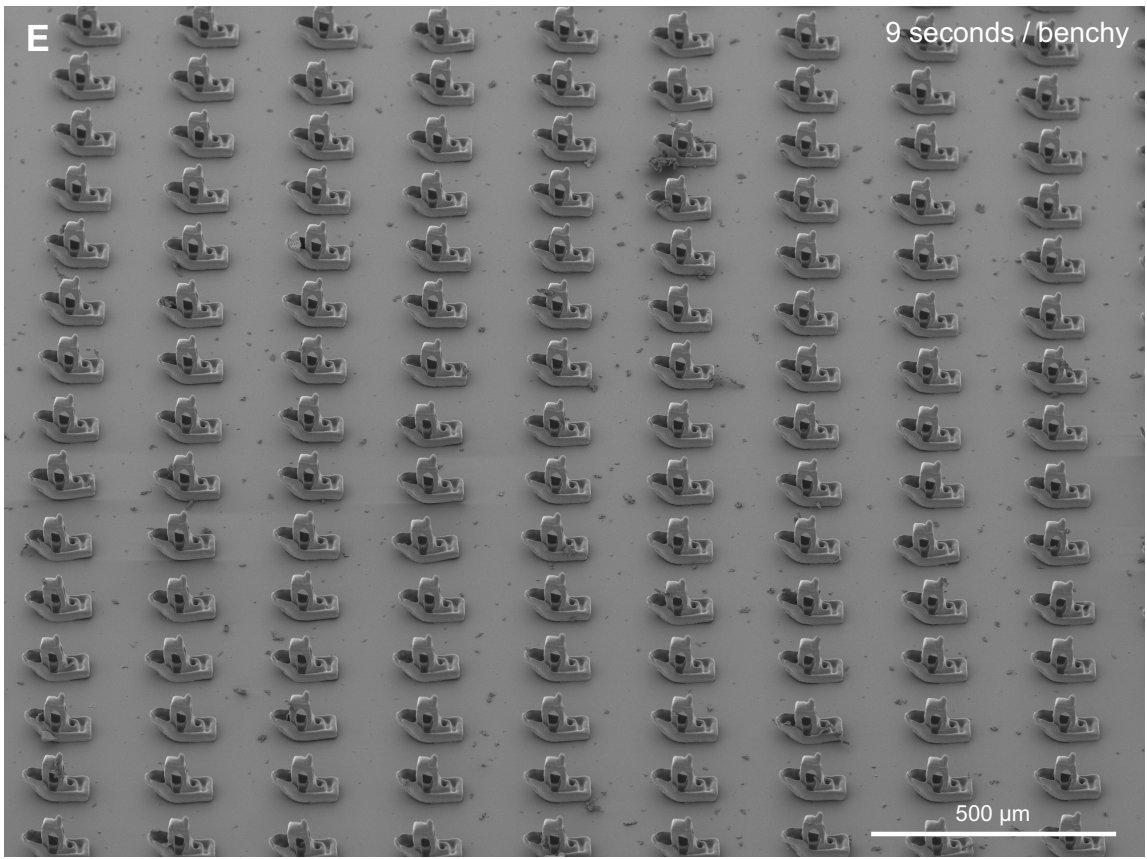

**Figure 3. Algorithmic correction for high-fidelity printing. (A)** The process of 3D deconvolution used to generate corrected projection images. **(B)** SEM image of a Benchy structure printed using uncorrected binary images. **(C, D)** SEM images of Benchy structures printed with final corrected images derived from the 3D deconvolution shown in (A), viewed from (C) the side and (D) the front. **(E)** SEM image showing more than 130 Benchy structures from the side view. SEM images in (B - E) are tilted at 40°.

We next sought to fabricate more complex 3D structures. We selected a 3D tugboat ("Benchy"),[31] a widely used benchmark structure with intricate geometries, to rigorously validate the applicability of our technology. In our initial attempt with only materials and optical hardware engineering (**Fig. 3B** or **Fig. S9A**), we used binary images (**Fig. S10**, first column) with equal projection times for each layer. This resulted in a Benchy structure without a visible open cabin or other expected structural features, indicating significantly more polymerization ("overcure") than expected. The overcuring phenomenon arises from proximity effects:[35–37] when a set of projected images contains non-periodic, complex shapes, the light distribution across the object becomes non-uniform due to the overlap of neighboring pixels. Regions with more activated pixels receive a higher total light dose, while regions with fewer activated pixels receive less, resulting in overcuring in some areas and undercuring in others. As indicated by the three-line bridge results (**Fig. S8**), radical accumulation is also non-negligible, signifying the control of radical diffusion is critical for successful printing. To account for varying shapes in each layer, we manually adjusted the exposure time for each layer (e.g., reducing exposure for layers with large projection areas), but no significant improvement was observed (**Fig. S9B**). Despite the quadratic nature of the nonlinear upconversion process, proximity effects significantly impaired high-fidelity printing. This phenomenon has also been observed in two-photon polymerization (2PP) during attempts at parallelization.[35]

Algorithmic corrections to the projected imagines provided a solution to reduce proximity effects for targeted structures by ensuring more uniform light projection. We developed an algorithm based on Richardson-Lucy (RL) 3D deconvolution[38–40] (**Fig. 3A**). First, we modeled an optical kernel and a diffusion kernel based on the optical setup (objective parameters and resin refractive index), the quadratic nature of the UC process, and material properties (e.g., radical diffusion coefficient approximation[40]). These kernels were combined through 3D convolution to generate a final kernel. Intuitively, this kernel represents radical generation, which is proportional to the square of the light intensity, followed by isotropic diffusion based on the diffusion coefficient and exposure time. Next, the final kernel was convolved with the target structure (e.g., the Benchy) to simulate the printing outcome, which closely resembled the uncorrected binary projection results with uniform exposure time (**Fig. 3B**). By performing 3D deconvolution, we generated corrected grayscale images for projection to mitigate proximity and diffusion effects. As shown in **Fig. S10** (second column), the RL algorithm's corrections are visually intuitive: pixels surrounded by more activated pixels are dimmed, while those with fewer neighboring pixels are brightened. However, the printing results were not fully satisfactory. Although the RL algorithm improved printing fidelity, as evidenced by Benchy's more defined cabin shape (**Fig. S9C** and **S9D**), the results were insufficient to fully correct proximity effects. Beyond the RL deconvolution, we added a sparsity-based algorithm to better correct for proximity effects. This correction selectively deactivates pixels based on the grayscale values of neighboring pixels in 3D, including those on the same plane and in adjacent layers. If the brightness of neighboring pixels exceeds a certain threshold, nearby pixels are turned off. With this approach (**Fig. S10**, the third column), we achieved significantly improved Benchy structures (**Fig. 3C** and **3D**).

The deconvolution and sparsity algorithmic corrections not only enhanced the fidelity of the target object but also helped to maintain the printing speed. The recording presented in **Supplementary Video 1** shows Benchy structures printed from a top-down view. In **Fig. 3E**, we demonstrated the repeatability and scalability of the technology by sequentially printing over 130 Benchy structures at a printing speed of 9 seconds/Benchy, with a build volume of approximately 120 μm × 50 μm × 150 μm = $9 \times 10^5$ μm$^3$ per Benchy.

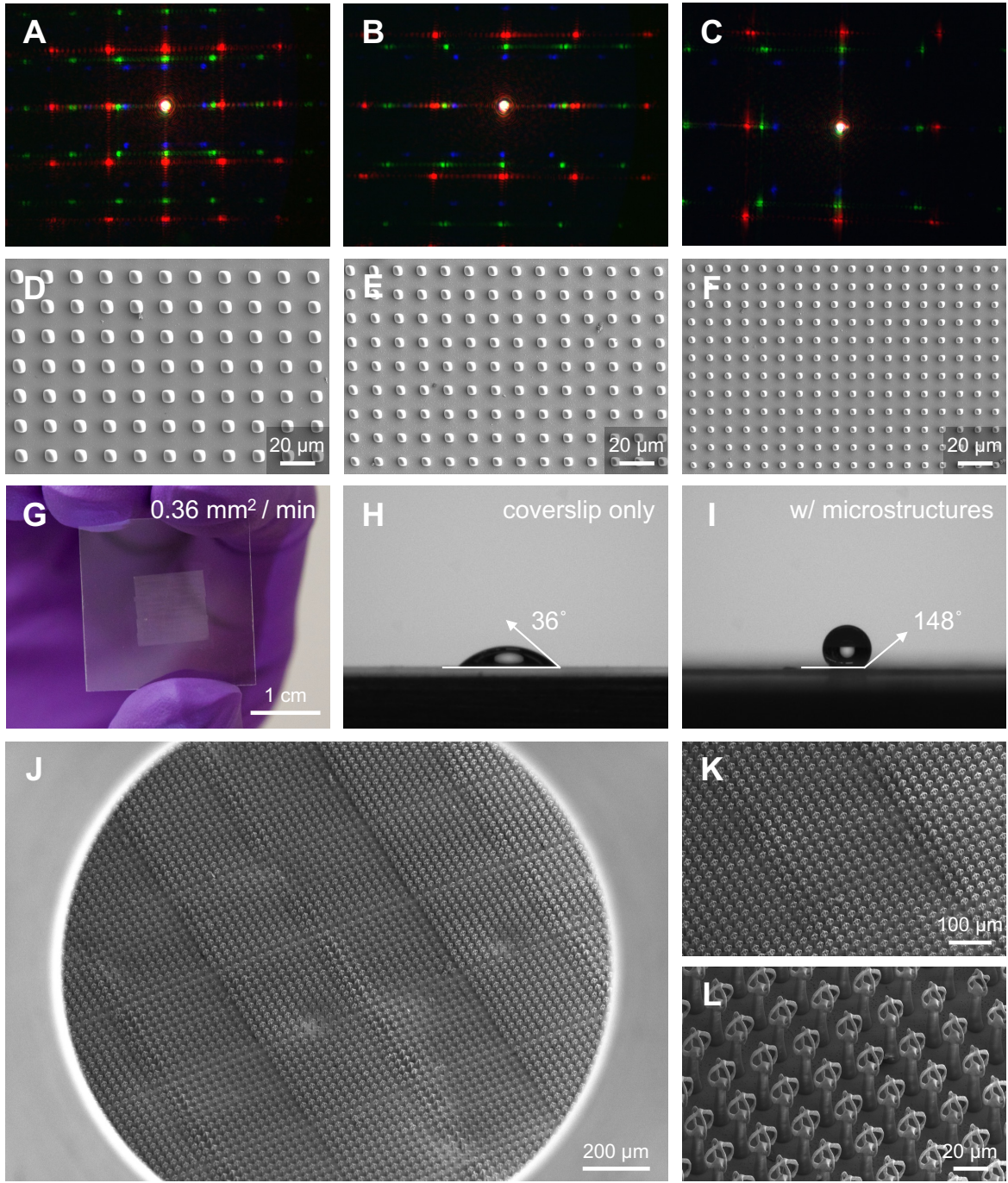

**Figure 4. Toward scaling micro-/nanomanufacturing for real-world applications. (A - C)** Diffraction patterns generated by directing lasers of different wavelengths (447 nm (blue), 532 nm (green), and 635 nm (red)) onto printed pillar arrays with varying periodicities. The images of different wavelengths were captured separately and combined using a Python script. **(D - F)** SEM top-down images of printed pillars with decreasing period from (D) to (F). **(G)** Photograph of a 1 cm × 1 cm hydrophobic surface printed on a coverslip. The printing speed was 0.36 mm$^2$/min, including stage movement time. **(H, I)** Contact angle measurements of (H) an untreated coverslip and (I) a coverslip with printed hydrophobic microstructures. 0.2 µL of water droplets were applied onto the samples. **(J - L)** SEM images showing tilted views (40°) of the hydrophobic microstructures at different magnifications.

Beyond printing complex 3D structures like Benchy, we aimed to demonstrate the technology's potential for real-world applications. One promising application is the scalable fabrication of diffraction gratings for light manipulation. To showcase this capability, we printed periodic square arrays to serve as simple, scalable diffraction gratings. By adjusting the pillar periodicity and the wavelength of the input light, we produced distinct diffraction patterns, as shown in **Fig. 4A - C**. **Fig. 4D - F** presents SEM images of the fabricated periodic pillars corresponding to the diffraction patterns in **Fig. 4A - C**. These pillars were fabricated at a rate of 4.35 mm$^2$/min, including exposure and stage movement times, using half of the total available voxels without sparsity correction. Light-off stage movement accounted for approximately 85% of the total time, highlighting a straightforward method to improve throughput in future work.

Finally, we fabricated hydrophobic microstructures to further demonstrate the potential of parallelized nanoscale 3D printing. The design of these structures, inspired by previous research on the hydrophobic surfaces of salvinia molesta leaves,[41,42] is shown in **Fig. S11**. We fabricated these structures over a 1 cm × 1 cm area on a glass coverslip (**Fig. 4G**) at a printing speed of 0.36 mm$^2$/min, accounting for both exposure and stage movement times. Light-off stage movement represented about half the time. **Supplementary Video 2** captures the printing process of the hydrophobic structures. Contact angle measurements were performed to compare the hydrophobicity of an untreated coverslip with that of the microstructured coverslip. The untreated coverslip was hydrophilic, with a contact angle of 36° (**Fig. 4H**). In contrast, the microstructured coverslip exhibited a contact angle of 148°, indicating strong hydrophobicity (**Fig. 4I**). **Supplementary Video 3** shows a 6 µL water droplet rolling off the hydrophobic surface. SEM images of the hydrophobic structures (**Fig. 4J - L**) closely resembles the CAD design (**Fig. S11**), demonstrating high fidelity and excellent reproducibility across the entire area.

## Discussion and Future Work

In summary, 3D micro- and nanoscale printing using TTA-UC is shown to achieve remarkable advancements, enabling low-power, scalable nanomanufacturing with minimum feature sizes down to 230 nm and printing speeds up to 112 million voxels/s at an input power of 7.0 nW/voxel. We developed a photoresin incorporating a TTA-UC system to enable controlled polymerization and emphasized the importance of suppressing undesired 1PA reactions. Our optical systems deliver photons with precision approaching the diffraction limit, and our algorithms effectively address proximity and diffusion effects. By engineering materials, optics, and algorithms in a cohesive, iterative process, we demonstrated the technology's transformative potential for crafting large-scale, precise, custom micro- and nanopatterned materials.

Looking ahead, the printing resin could be adapted for materials such as glass[1] and ceramics[43] and engineered to incorporate additional functionalities, such as biocompatibility.[44,45] In addition, more advanced algorithms can be explored to further minimize proximity effects and simplify workflows. This technology holds significant promise for diverse applications, including constructing metasurfaces for augmented reality/virtual reality, creating cell scaffolds for tissue engineering, fabricating nanopatterned radio frequency identification antennas, and developing photonic integrated circuits, highlighting the possibility of this technology to make a substantial impact at scale.

## Acknowledgements

This research was partially sponsored by the Army Research Office and was accomplished under Cooperative Agreement Number W911NF-24-2-0099. The views and conclusions contained in this document are those of the authors and should not be interpreted as representing the official policies, either expressed or implied, of the Army Research Office or the U.S. Government. The U.S. Government is authorized to reproduce and distribute reprints for Government purposes notwithstanding any copyright notation herein. Q.Z. acknowledges Sunlin & Priscilla Chou Graduate Fellowship and a Stanford Graduate Fellowship (SGF) in Science & Engineering as a STMicroelectronics Fellow. Q.Z. and H.Y. thank Dr. Pingyu Wang for insightful discussion on printing and washing procedures. Q.Z. and Q.L. gratefully acknowledge Dr. Ona Segura Lecina for assistance with video recording for the paper. Q.Z. and Q.L. also thank Dr. Yihang Chen and Han Cui for their help with sample characterization. A.O.G. acknowledges the support of a National Science Foundation Graduate Research Fellowship under grant DGE-1656518 and an SGF in Science & Engineering as a Scott A. and Geraldine D. Macomber Fellow. M.H. acknowledges the support of the Department of Electrical Engineering at Stanford University. K.F. acknowledges funding and support from the Stanford Energy Postdoctoral Fellowship and the Precourt Institute for Energy. D.L. acknowledges an SGF in Science & Engineering as a Gabilan Fellow and Stanford EDGE Fellowship. N.M. acknowledges an SGF in Science & Engineering as a Dr. Robert N. Noyce Fellow and the support of a National Science Foundation Graduate Research Fellowship under grant DGE-2146755. P.N. acknowledges an SGF in Science & Engineering as a Gabilan Fellow and the Chevron Fellowship in Energy. T.H.S. gratefully acknowledges support from the Arnold and Mabel Beckman Foundation. L.P. acknowledges the support of a National Science Foundation Graduate Research Fellowship under grant DGE-2146755. S.F. acknowledges the support from Stanford University as a Diversifying Academia, Recruiting Excellence (DARE) Fellow, the U.S. Department of Energy (DOE) Building Technologies Office (BTO) as an IBUILD Graduate Research Fellow, SGF in Science & Engineering as a P. Michael Farmwald Fellow, and of the National GEM Consortium as a GEM Fellow. This research was performed under an appointment to the Building Technologies Office (BTO) IBUILD Graduate Research Fellowship administered by the Oak Ridge Institute for Science and Education (ORISE) and managed by Oak Ridge National Laboratory (ORNL) for the DOE. ORISE is managed by Oak Ridge Associated Universities (ORAU). All opinions expressed in this paper are the authors' and do not necessarily reflect the policies and views of the DOE, EERE, BTO, ORISE, ORAU, or ORNL. M.S. acknowledges the financial support through a Doc.Mobility Fellowship from the Swiss National Science Foundation with Grant No. 187676. In addition, M.S. acknowledges the financial support of a fellowship from "la Caixa" Foundation (ID 100010434). The fellowship code is LCF/BQ/IN17/11620040. Further, M.S. has received funding from the European Union's Horizon 2020 research and innovation program under the Marie Skłodowska-Curie Grant Agreement No. 713673. Part of this work was performed at the Stanford Nano Shared Facilities, supported by the National Science Foundation under award ECCS-2026822.

## Author contributions

Q.Z. designed and conducted experiments, analyzed data, and wrote the manuscript. H.Y. and Q.L. contributed to experimental design, data analysis, and manuscript preparation. H.Y., Q.L., and A.O.G. conducted printing experiments. A.O.G., M.H., K.F., H.N., N.M., and M.S. assisted with the construction and characterization of optical setups and data analysis. D.L., N.M., and T.H.S. contributed to experimental design and printing experiments. P.N., L.P., and S.F. assisted with materials characterization and data analysis. D.N.C. supervised the research. All authors contributed to the manuscript editing.

## Competing interests

D.N.C. is a co-founder of and Chief Scientific Advisor to Quadratic 3D, Inc. Stanford University has filed a patent based on this work.

## Materials & Correspondence

All chemicals were used as received.

1. palladium(II)-meso-tetraphenyl-tetrabenzoporphyrin (PdTPTBP) was purchased from Frontier Specialty Chemicals Inc., catalog number 41217.
2. 9,10-bis((diisopropyl(octyl)silyl)ethynyl)anthracene (NODIPS-an) was purchased from HAARES ChemTech Inc.
3. 1-phenylazo-2-naphthol (Sudan I) was purchased from Sigma-Aldrich, catalog number 103624-25G.
4. bis(2,2,6,6-tetramethyl-4-piperidyl-1-oxyl) sebacate (BTPOS) was purchased from Tokyo Chemical Industry, catalog number B5642.
5. di(trimethylolpropane) tetraacrylate (DTMPTA) was purchased from Sigma-Aldrich, catalog number 408360-100ML.
6. N,N-dimethylacrylamide (DMAA) was purchased from Sigma-Aldrich, catalog number 274135-5ML.
7. Bis(4-methoxybenzoyl)diethylgermanium (Ivocerin) was purchased from SYNTHON Chemicals GmbH & Co. KG, catalog number ST03909.
8. 2-(1-methoxy)propyl acetate was purchased from Thermo Scientific Chemicals, catalog number 413900010.
9. 2-propanol was purchased from Fisher Chemical™, catalog number 67-63-0.
10. Novec™ 7100 Engineered Fluid was purchased from Sigma-Aldrich, catalog number SHH0002-1L.
11. Acetone was purchased from Fisher Chemical™, catalog number 67-64-1.
12. PELCO® Conductive Silver Paint was purchased from Ted Pella Inc, catalog number 16062.
13. Immersion Oil Type A was purchased from Electron Microscopy Sciences, catalog number 16907-01.

Other supplies and equipment used in the experiments:

1. Silicone Isolators™ was purchased from Grace Bio-Labs, catalog number 664504.
2. SEM Pin Stub Specimen Mounts was purchased from Ted Pella Inc, catalog number 16144-9.
3. Fisherbrand™ Lens Paper was purchased from Fisher Scientific, catalog number 11-996.
4. Corning® 25x25 mm Square #1 Cover Glass was purchased from Corning, catalog number 2845-25.
5. Electron Microscopy Sciences Glass Slide Plain, 25x75, 1.2 mm thick, was purchased from Electron Microscopy Sciences, catalog number 50-189-9727.
6. The LED used on the printing setup was purchased from Luminus Devices, Inc., catalog number CBT-90-RX.
7. The DMD used on the printing setup was purchased from Vialux GmbH, catalog number V-6501 VIS.
8. The mirostage used on the printing setup was purchased from Mad City Labs, MMP25 with AC-V 25 adapter set.

# Methods

## Resin preparation

Four stock solutions were prepared in DMAA within a nitrogen-purged glove box (<0.5 ppm $O_2$):

1. 2 mg/mL PdTPTBP
2. 12.5 mg/mL NODIPS-an
3. 10 mg/mL Sudan I
4. 10 mg/mL BTPOS

Typically, 2 - 3 mL of each stock solution was prepared, stirred for at least 4 hours before initial use, and stored in the glove box for no longer than one week.

To prepare the resin, 20 mg Ivocerin was added to 2 mL of DTMPTA. Then, the following volumes of stock solutions were added in a nitrogen-purged glove box:

1. 100 µL of PdTPTBP in DMAA for a final concentration of 100 ppm
2. 200 µL of NODIPS-an in DMAA for a final concentration of 1250 ppm
3. 5 µL of Sudan I in DMAA for a final concentration of 25 ppm
4. 30 µL of BTPOS in DMAA for a final concentration of 150 ppm

The resulting resin was stirred overnight in the glove box at 400 rpm. Stirring faster than this speed is not recommended, as it may generate excessive bubbles in the solution that decrease optical clarity of the resin.

## Printing sample preparation

Since the upconversion process is oxygen-sensitive, we designed sample holders to securely seal the resin. Silicone isolators with a thickness of 0.5 mm were utilized. One side of the isolator was covered with a 0.15 mm or 0.17 mm thick coverslip. Then, the silicone isolator was filled with 350 µL of resin in a nitrogen-purged glove box. The resin was encapsulated with a 1.2 mm thick microscope glass slide, which was pressed firmly onto the isolator to ensure a tight seal. The microscope glass slide with the sealed resin was attached to the sample holder on the microstage. The sealed sample remained stable in ambient conditions for at least 8 hours without significant changes in printability.

## Washing procedure

After printing, the microscope glass slide and silicone isolator were removed. The coverslip with printed structures was gently immersed in 2-(1-methoxy)propyl acetate for 10 minutes. The sample was then transferred to 2-propanol and immersed for 5 minutes while undergoing UV lamp post-curing. Finally, the sample was held with tweezers and immersed in Novec™ 7100 Engineered Fluid for 1 minute. The sample was then attached to an SEM sample holder with silver paste for imaging.

## Gold sputter

All the printed samples were sputtered 12.5 nm gold using Leica EM ACE600 sputter coater for conductivity under SEM.

## Scanning electron microscopy (SEM)

SEM was performed on one of the following instruments.

1. Thermo Fisher Scientific Apreo S LoVac scanning electron microscope
2. FEI Magellan 400 XHR scanning electron microscope
3. JEOL JSM-IT500HR environmental scanning electron microscope

The instrument was selected based solely on its availability.

## UV-vis measurements

UV-vis absorption spectra shown in **Fig. 1D** were collected with an Agilent Cary 6000i UV/Vis/NIR.

## Photoluminescence measurements

Photoluminescence (PL) spectra shown in **Fig. 1D** and **Fig. S4** were collected on a custom setup. A 635 nm laser (MRL-III-635-500mW, Changchun New Industries Optoelectronics Tech. Co., Ltd. (CNI)) excited the sample, PL was collected at 90 degrees using a collection lens, and the appropriate emission filter was used to avoid saturating the spectrometer (QE Pro, QEPRO-XR, Ocean Insight) with laser light (600 nm short pass filter for upconversion PL; 650 nm long pass filter for phosphorescence.

## Threshold measurements

Threshold measurements in **Fig. S3** were collected on a custom setup, where a 635 nm laser (MRL-III-635-500mW, Changchun New Industries Optoelectronics Tech. Co., Ltd. (CNI)) was focused on the sample, and UCPL was collected at 90 degrees by the spectrometer with a 575 nm short pass filter. The power of the laser at the focal point was measured using a Thorlabs power meter (PM100D meter with S120VC sensor). The image of the laser spot was captured by a CMOS scientific camera (CS165MU, ThorLabs, Inc.) and analyzed in ImageJ to determine the spot size. Different beam intensities were achieved by attenuating the laser with neutral density filters (NEK01, ThorLabs, Inc.). Upconverted emission spectra were integrated and plotted against excitation intensity in a log-log plot. The resulting plot was analyzed to find the linear and quadratic regimes, and the intersection between these regimes was used to interpolate the threshold intensity.

## Time-resolved upconversion photoluminescence (TRUCPL) measurements

The time-resolved upconversion photoluminescence decay spectra were collected using a Streakscope C10627 from Hamamatsu Corp. A 635 nm laser (MRL-III-635-500mW, Changchun New Industries Optoelectronics Tech. Co., Ltd. (CNI)) was pulsed by a function generator RIGOL DG812 with a square wave function with a 5% duty cycle.

## Point spread function measurements

Point spread function measurements were performed following the procedure in **Ref. 33**. A bottle of 100 nm gold nanoparticles dispersed in 0.1 mM PBS (Sigma-Aldrich, catalog number 753688-25ML) was vigorously vortexed. In a 15 mL conical tube, 100 µL of the gold nanoparticle solution was diluted to 10 mL with Milli-Q water, achieving a 1:100 dilution. In a second 15 mL conical tube, 100 µL of the 1:100 diluted solution was further diluted to 10 mL with Milli-Q water, resulting in a 1:2 dilution. The second diluted solution was sonicated in a water bath for 20 minutes to disperse aggregated nanoparticles. In a third 15 mL conical tube, 100 µL of the sonicated solution was diluted to 10 mL with 9 mL of Milli-Q water and 900 µL of 70% (vol/vol) ethanol. Microscope slides and coverslips were cleaned with 70% ethanol. The coverslips were covered with aluminum foil to prevent dust contamination and allowed to dry for at least 2 hours. Immediately before use, the coverslips were treated with UV-ozone for 10 minutes to enhance hydrophilicity. The final solution was vortexed immediately before use, and 15 µL was pipetted onto a clean coverslip. To locate the nanoparticles, a permanent marker was used to draw a circle around the dried spot on the underside of the coverslip. A 15 µL drop of ProLong Gold (Invitrogen™, catalog number P36934) was applied to a microscope slide, and the coverslip with nanoparticles was mounted onto it. A cotton swab was used to gently press the center of the coverslip, pushing any air bubbles in the mounting medium to the edges. The sample was then left in the dark overnight to allow the ProLong Gold to cure.

## Three-line bridge printing

Each line is one pixel wide, and each bridge is separated by 250 pixels. The vertical supports were printed first with a projection of 15 layers and a 5 µm step size. Then, the stage was mechanically moved up 50 µm, and the three-line image was projected.

## Contact angle measurements

Contact angle measurements were performed on Rame-Hart 290.

# Extended data

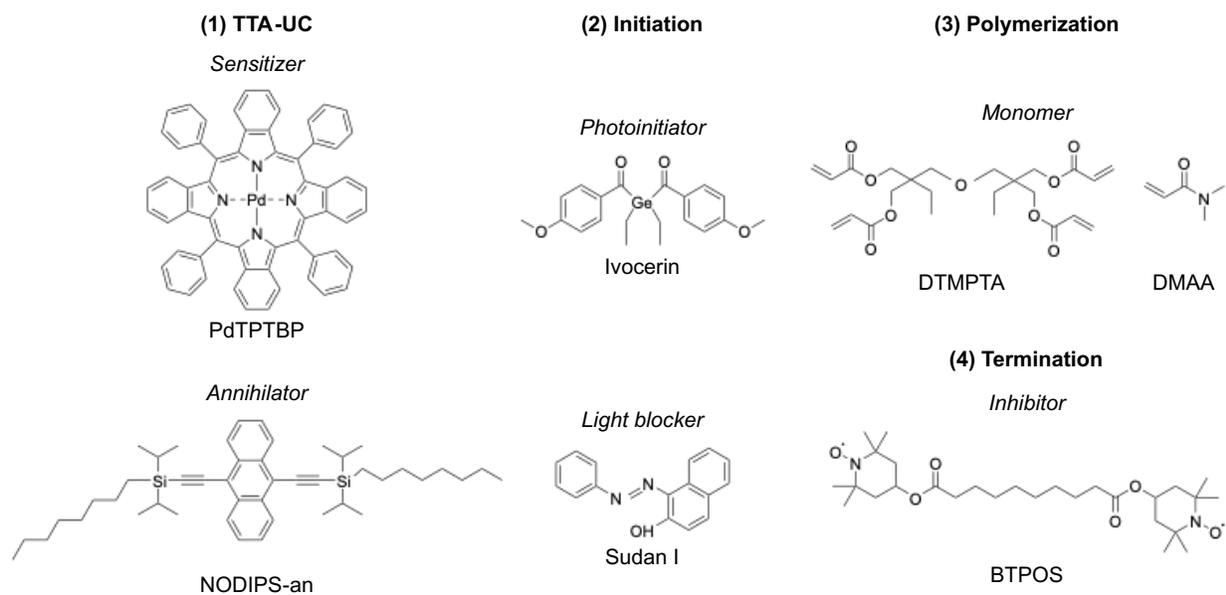

**Figure S1.** Molecular structure and function of each resin component.

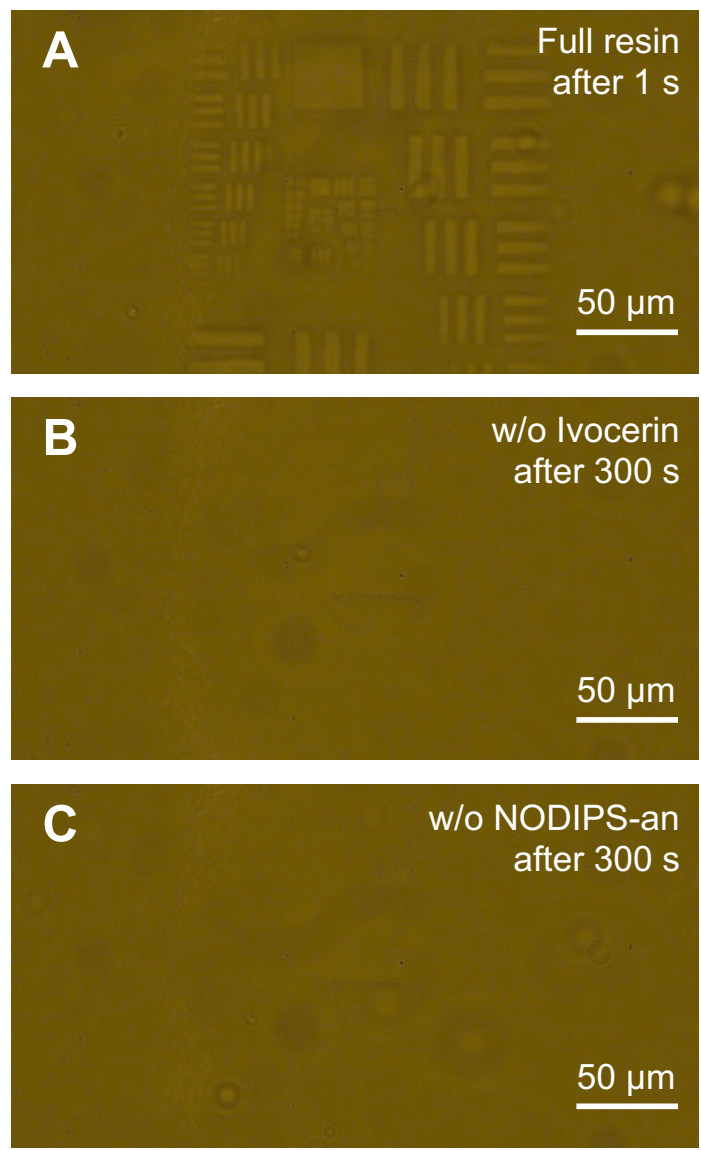

**Figure S2.** Microscopic images of resin after a specified duration of red-light projection. The projected pattern is a 1951 USAF resolution test chart. Note that the brightness of the images has been adjusted for better visualization. **(A)** Image of the complete, optimized resin after 1 s of projection. **(B)** Image of resin without the photoinitiator Ivocerin, after 300 s of projection. **(C)** Image of resin without the annihilator NODIPS-an, after 300 s of projection. The full resin polymerized at least two orders of magnitude faster than the control samples (B and C), indicating that the undesired 1PA reaction is significantly suppressed and that the main pathway for polymerization is through TTA-UC and energy transfer to the photoinitiator.

## Supplementary Note 1

We conducted threshold measurements in toluene to assess the effects of these additives at concentrations consistent with those in the resin, as summarized in **Fig. S3**. Four sample conditions were tested: (1) the sensitizer–annihilator pair alone (control), (2) with added BTPOS only, (3) with added Sudan I only, and (4) with both additives. The UC thresholds for the control and BTPOS-only samples were 566 and 463 mW/cm², respectively, indicating BTPOS had minimal impact. In contrast, the addition of Sudan I increased the UC threshold over 17-fold to 9,896 mW/cm². When both BTPOS and Sudan I were included, the threshold rose to 11,954 mW/cm², comparable to Sudan I alone. This dramatic threshold increase suggests that Sudan I may be quenching triplets generated by the sensitizer or annihilator. However, we were unable to perform similar measurements in monomer formulations due to excessive unwanted curing under high irradiation (linear regime), which prevented reliable data collection. Additional evidence of triplet quenching comes from phosphorescence measurements of the sensitizer (**Fig. S4**). Upon absorbing red photons, triplets in the sensitizer may either transfer to the annihilator or relax to the ground state via phosphorescence. Therefore, phosphorescence intensity reflects the behavior of these triplets. In samples containing Sudan I, phosphorescence intensity dropped to approximately one-quarter of the control, suggesting that Sudan I effectively scavenges triplets from the sensitizer. We further validated this hypothesis using time-resolved upconversion photoluminescence (TRUCPL) measurements. As shown in **Fig. S5** (top row), **Fig. S6**, and **Table S1**, the UC lifetime was significantly quenched by Sudan I in the quadratic regime (below the threshold), while BTPOS showed no significant effect. Since triplet lifetimes dominate UC lifetime in the quadratic regime, this result supports Sudan I's role as a triplet quencher. In contrast, in the linear regime, where triplets are rapidly generated and consumed, the UC lifetimes for all four samples were nearly identical (**Fig. S5** (bottom row); **Fig. S7**; **Table S2**), indicating singlet-state dynamics dominate under these conditions.

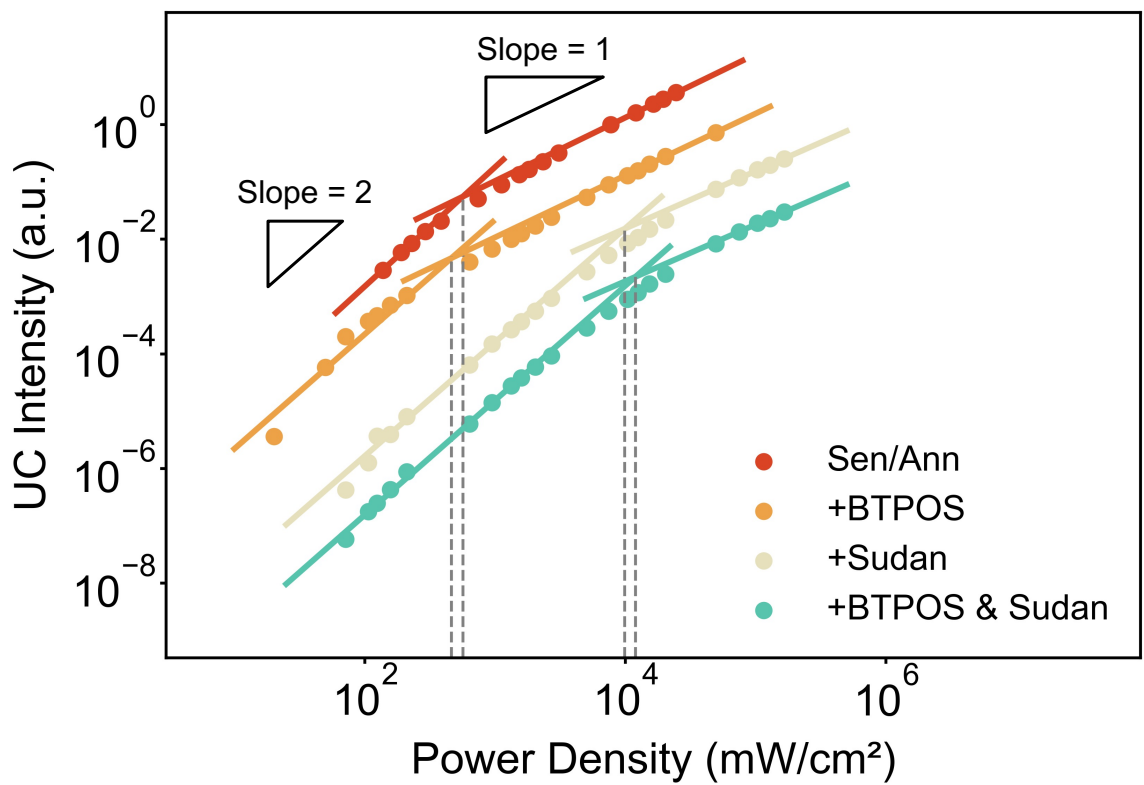

**Figure S3.** Threshold measurements of the upconversion system with different additives in toluene. The UC intensities for each sample are scaled by specific factors to enhance visualization and are thus not directly comparable as absolute values.

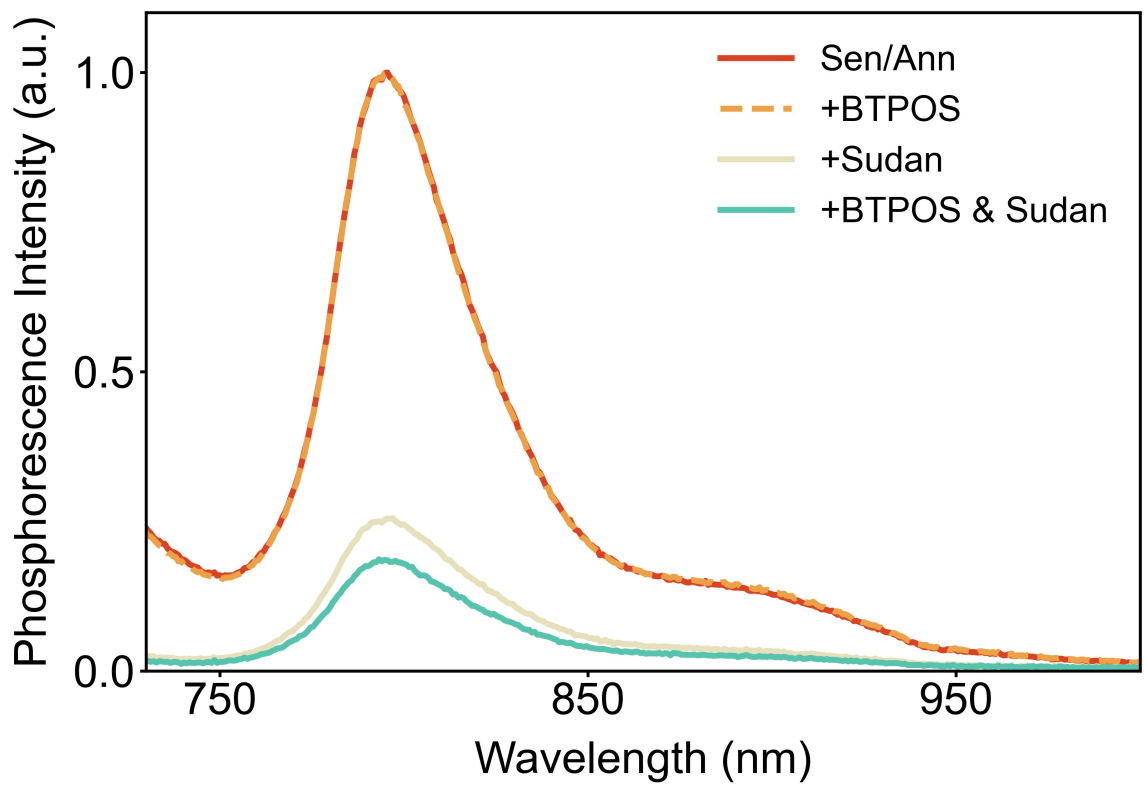

**Figure S4.** Relative phosphorescence intensity of PdTPTBP (Sen) with different additives in toluene.

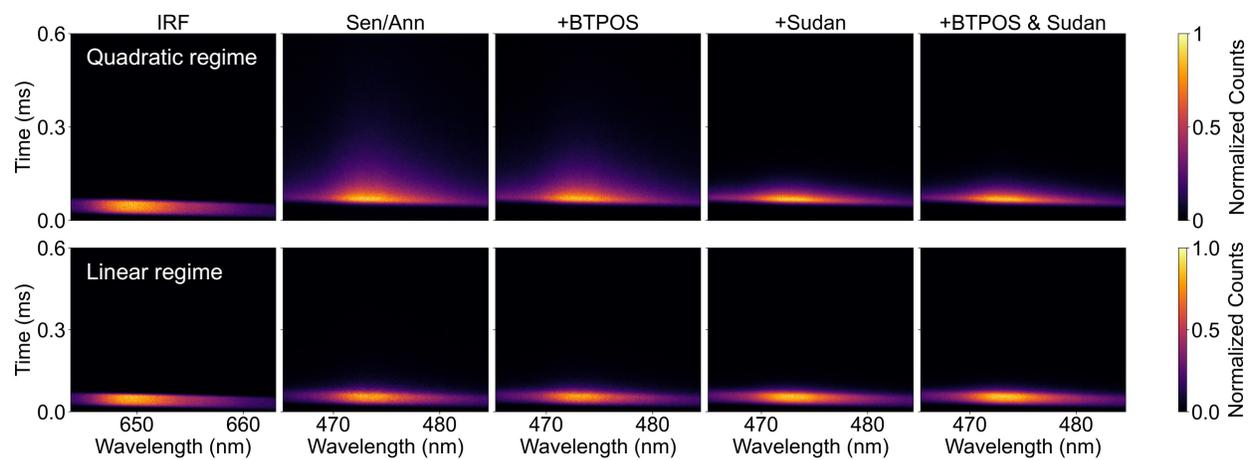

**Figure S5.** Time-resolved upconversion photoluminescence (TRUCPL) of the upconversion system with different additives in toluene.

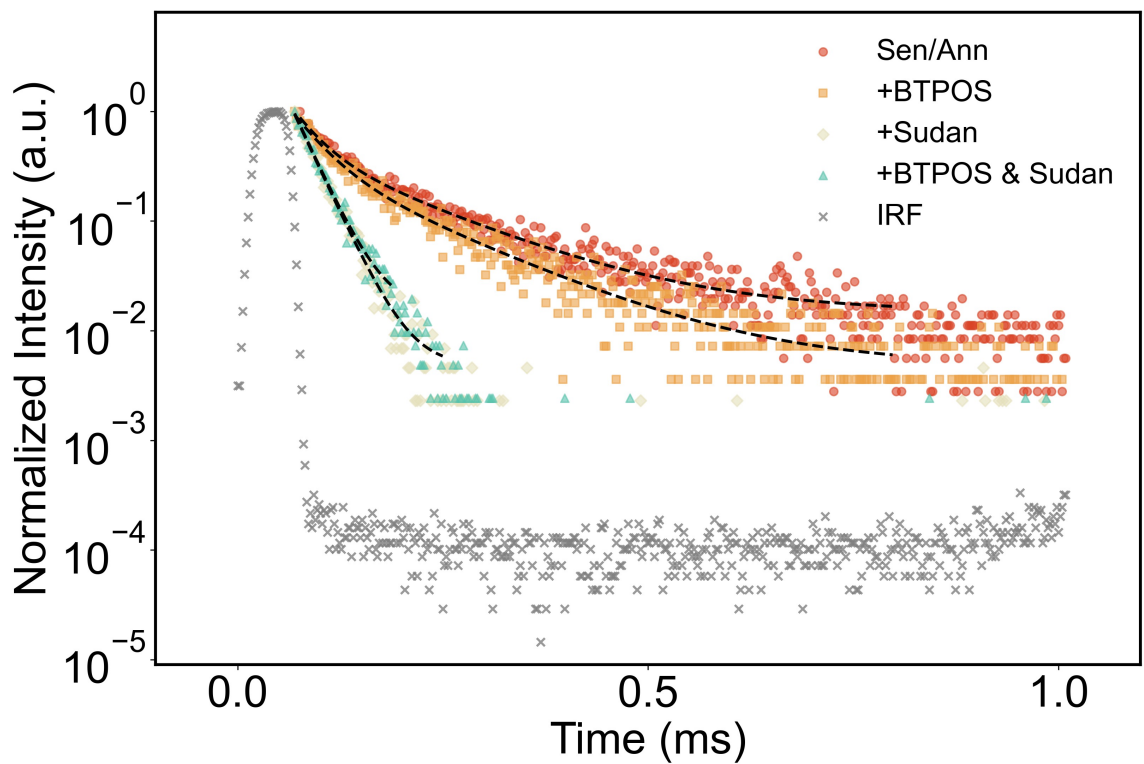

**Figure S6.** Log-scale plot and bi-exponential fitting for TRUCPL of the upconversion system with different additives in toluene in the quadratic regime.

**Table S1.** Parameters for bi-exponential fitting in **Fig. S6**.

|  | A1 (%) | $\tau_1$ (μs) | A2 (%) | $\tau_2$ (μs) | $\tau\_avg$ (μs) | R2 |
|---|---|---|---|---|---|---|
| **Sen/Ann** | 89 | 31 | 11 | 135 | 42 | 0.986 |
| **+BTPOS** | 89 | 35 | 11 | 130 | 45 | 0.987 |
| **+Sudan** | 97 | 27 | 3 | 2 | 26 | 0.995 |
| **+BTPOS & Sudan** | 97 | 26 | 3 | 2 | 26 | 0.992 |

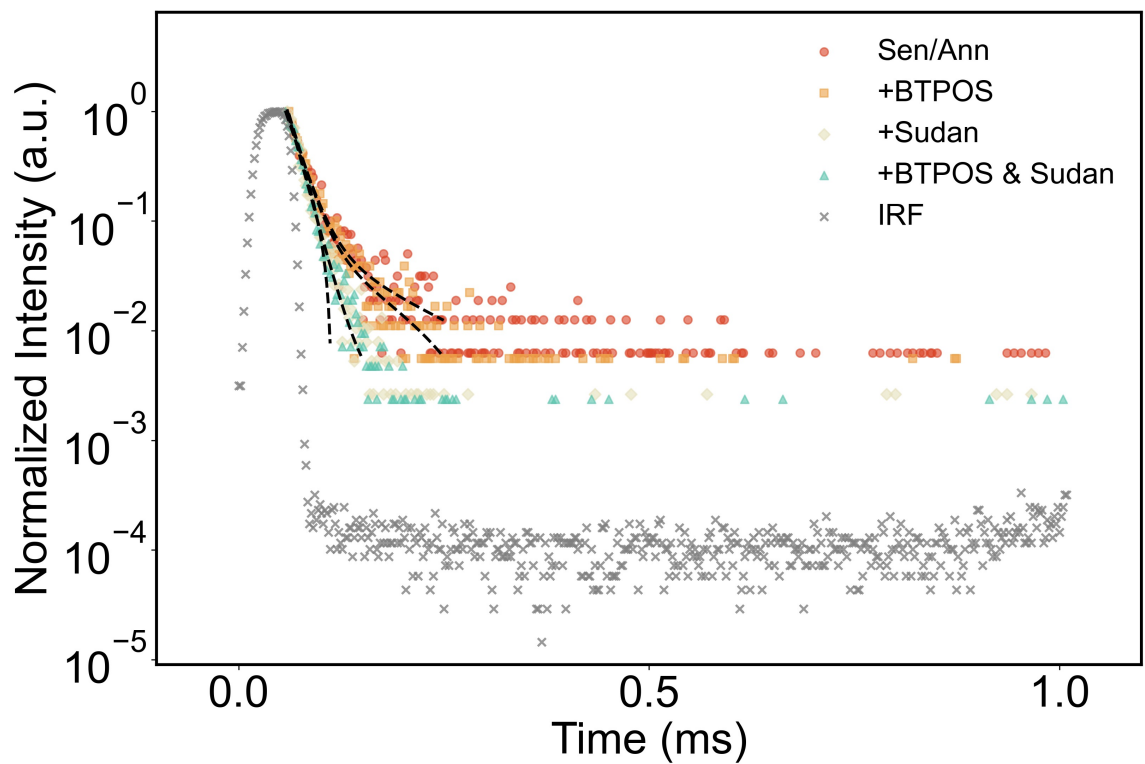

**Figure S7.** Log-scale plot and bi-exponential fitting for TRUCPL of the upconversion system with different additives in toluene in the linear regime.

**Table S2.** Parameters for bi-exponential fitting in **Fig. S7.**

|  | A1 (%) | $\tau 1$ (µs) | A2 (%) | $\tau 2$ (µs) | $\tau\_avg$ (µs) | R2 |
|---|---|---|---|---|---|---|
| **Sen/Ann** | 99.5 | 17 | 0.5 | 96 | 18 | 0.986 |
| **+BTPOS** | 99.6 | 16 | 0.4 | 105 | 17 | 0.991 |
| **+Sudan** | 99.8 | 16 | 0.2 | 2 | 16 | 0.986 |
| **+BTPOS & Sudan** | 95.5 | 22 | 4.5 | 22 | 22 | 0.986 |

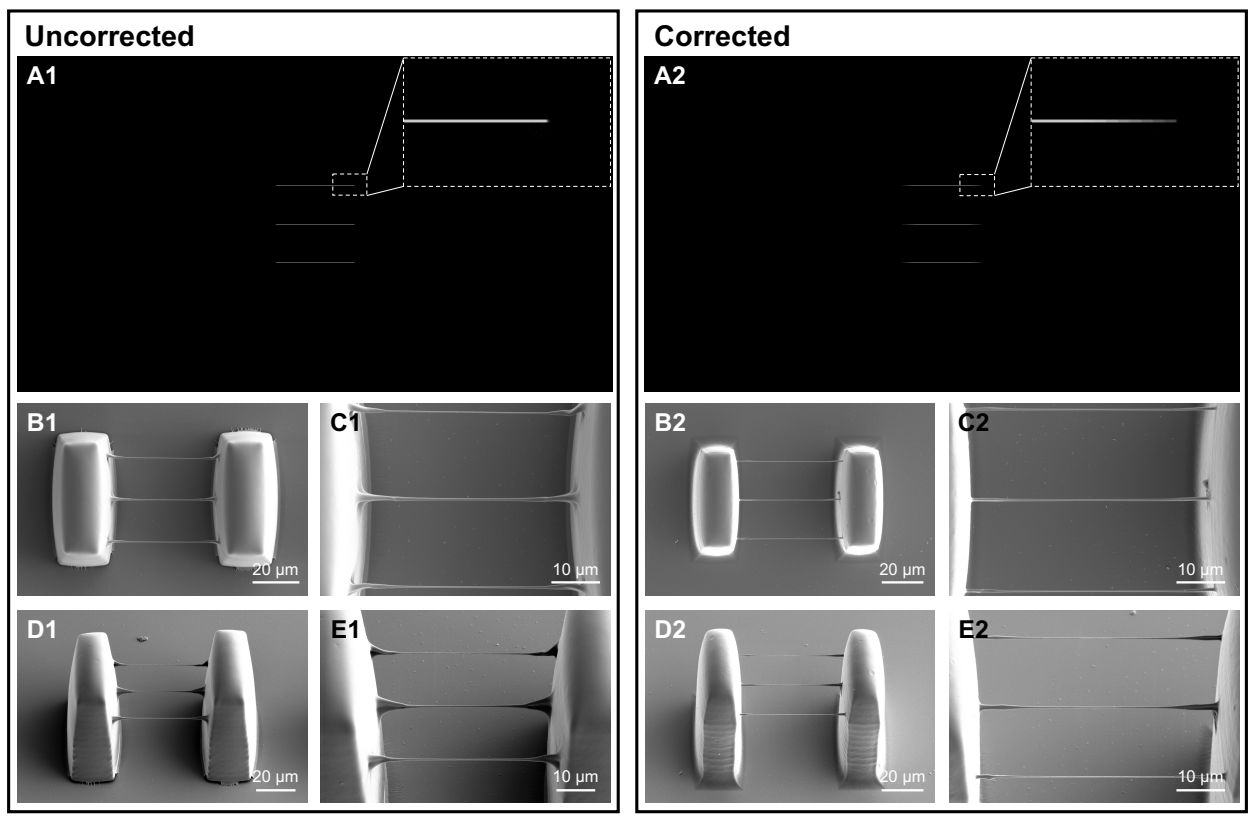

**Figure S8.** Comparison of uncorrected and corrected three-line bridges. **(A1, A2)** Projected images of uncorrected and corrected versions of three lines for resolution test. Both are single-pixel lines, and the difference is that pixels at the edge are adjusted to become dimmer for the corrected version. **(B1 - E1)** SEM images of uncorrected three-line bridge. (B1) and (C1) are from top-down view, while (D1) and (E1) are from a 40° tilted view. **(B2 - E2)** SEM images of corrected three-line bridge. (B2) and (C2) are from top-down view, while (D2) and (E2) are from a 40° tilted view.

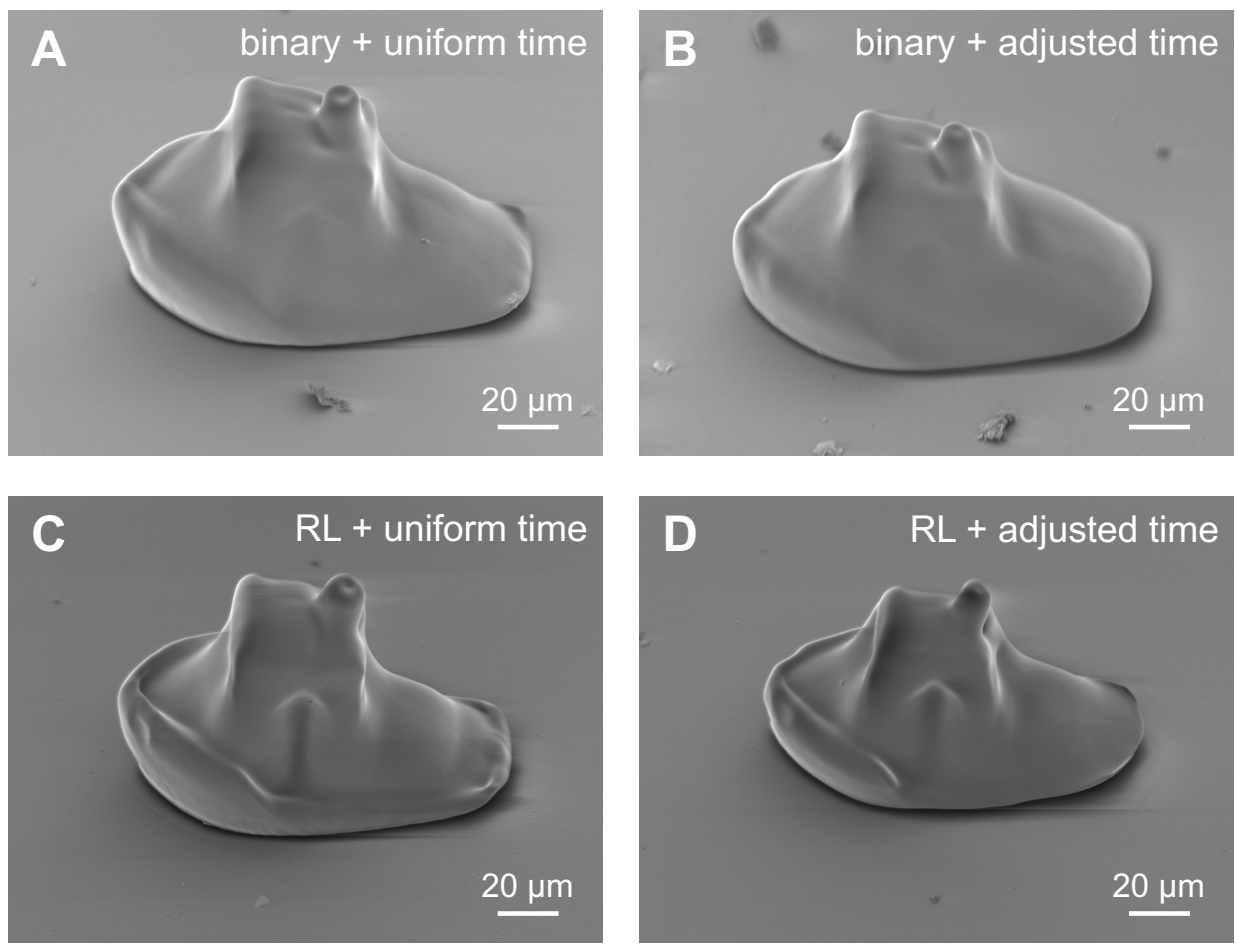

**Figure S9.** SEM images of benchy from tilted view (40°) with different corrections.

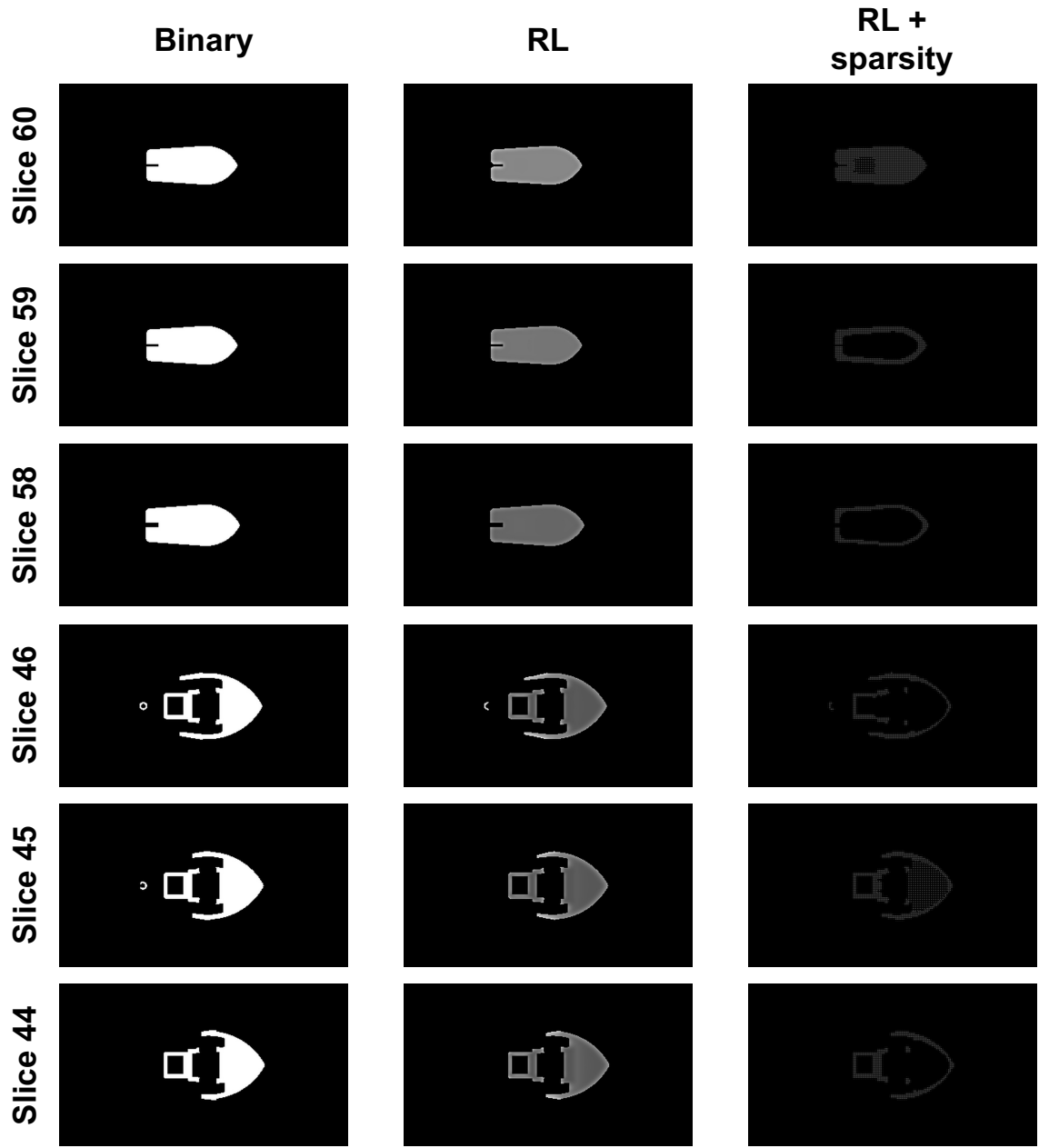

**Figure S10.** Comparison of uncorrected binary images, grayscale images with RL correction, and grayscale images with RL correction and sparsity. The brightness of the images for RL + sparsity was increased for better visualization.

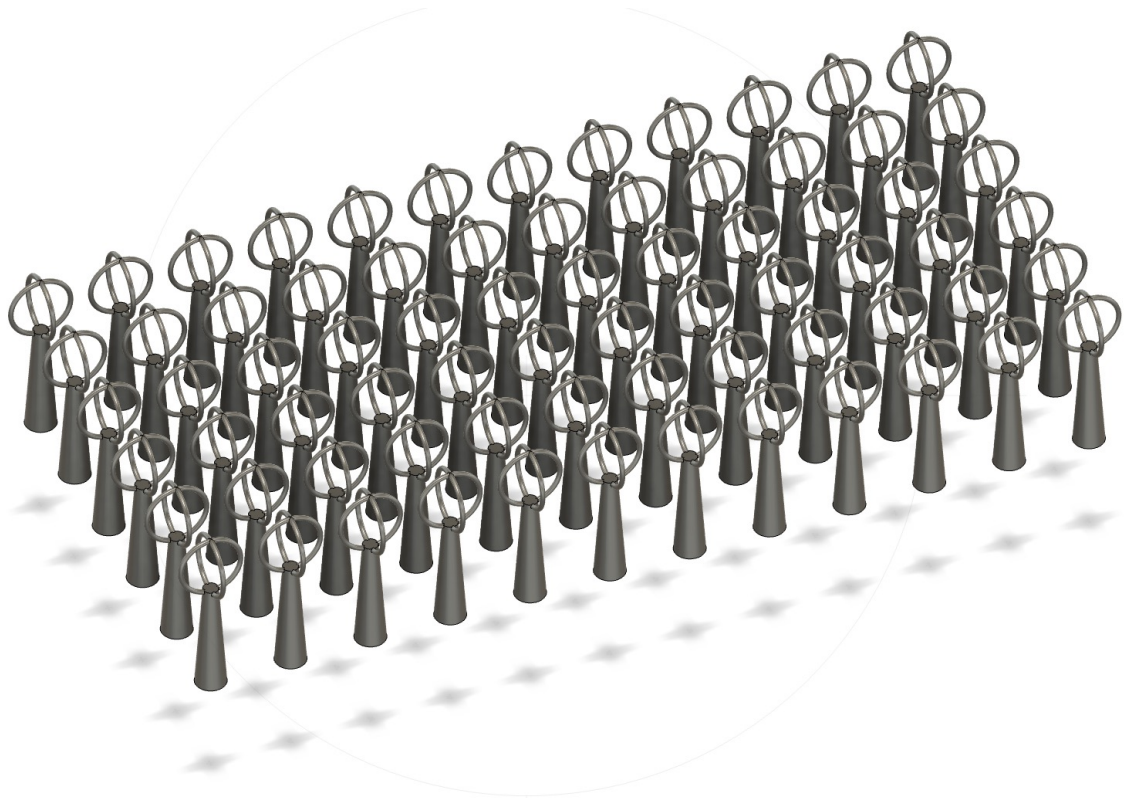

**Figure S11.** Image of CAD design for the bio-inspired hydrophobic structures.